\documentstyle[epsfig,aps,twocolumn]{revtex}

\newcommand{\pol}{\hat{\bf e}}
\newcommand{\rv}{{\bf r}}
\newcommand{\Ev}{{\bf E}}
\newcommand{\Dv}{{\bf D}}
\newcommand{\Pv}{{\bf P}}

\newcommand{\dv}{{\bf d}}

\newcommand{\Dc}{{\cal D}}

\newcommand{\Hc}{{\cal H}}
\newcommand{\Fc}{{\cal F}}

\newcommand{\Dkappav}{\Delta {\bbox \kappa}}

\newcommand{\qv}{{\bf q}}
\newcommand{\kv}{{\bf k}}

\newcommand{\eo}{\epsilon_0}
\newcommand{\beq}{\begin{equation}}
\newcommand{\eeq}{\end{equation}}
\newcommand{\bea}{\begin{eqnarray}}
\newcommand{\eea}{\end{eqnarray}}
\newcommand{\<}{\langle}
\renewcommand{\>}{\rangle}
\renewcommand{\(}{\left(}
\renewcommand{\)}{\right)}
\renewcommand{\[}{\left[}
\renewcommand{\]}{\right]}

\newcommand{\up}{\uparrow}
\newcommand{\down}{\downarrow}
\newcommand{\commentout}[1]{{}}

\begin{document}

\preprint{}
\draft
\title{Scattering of light and atoms in a Fermi-Dirac gas with BCS
pairing} 
\author{J. Ruostekoski} 
\address{Abteilung f\"ur Quantenphysik,
Universit\"at Ulm, D-89069 Ulm, Germany}
\date{\today}
\maketitle
\begin{abstract}
We theoretically study the optical properties of a Fermi-Dirac
gas in the presence of a superfluid state. We calculate the leading
quantum-statistical corrections to the standard column density result
of the electric susceptibility. We also consider the Bragg diffraction
of atoms by means of light-stimulated transitions of photons between
two intersecting laser beams. Bardeen-Cooper-Schrieffer pairing 
between atoms in different 
internal levels magnifies incoherent scattering processes.
The absorption linewidth of a Fermi-Dirac gas is broadened and 
shifted. Bardeen-Cooper-Schrieffer pairing introduces a collisional 
local-field shift that may
dramatically dominate the Lorentz-Lorenz shift.
For the case of the Bragg spectroscopy the static structure function may be
significantly increased due to superfluidity in the nearforward scattering.

\end{abstract}

\section{Introduction}

As a result of dramatic progress in cooling and trapping of
alkali-metal atomic gases the quantum statistics of atoms
has observable consequences. Perhaps, the most striking effect
is the Bose-Einstein (BE) condensation of bosonic atoms 
\cite{AND95,BRA95,DAV95}.
Nevertheless, Fermi-Dirac (FD) gases are also expected to exhibit
a rich and complex behavior.
One especially fascinating property of FD gases is that with 
effectively attractive interaction between different particles the
ground state of the system may become unstable with respect to the 
formation of bound pairs of quasiparticles or 
Cooper pairs \cite{LIF80,FET71}.
This effect is analogous to the Bardeen-Cooper-Schrieffer
(BCS) transition in superconductors.
FD gases have been an active subject of research already quite some time 
\cite{JAV95b,BAR96,STO96,HOU97,MAR98,MAR99,BUT97,MOD98,BUS98,JAV99,RUO99c,RUO99d,FER99,ZHA99,BRU98b,VIC99,BRU99}.

We recently showed \cite{RUO99d} that the coherent quasiparticle 
pairing between atoms in different internal levels may enhance the
optical interactions. In particular, the resonance line of
a FD gas (the extinction of light from coherent laser beam) 
is broadened and shifted 
as a result of the BCS pairing. In this paper
we present a more detailed study of the propagation of light
in a dilute FD gas in the presence of a superfluid state. We also
investigate the signatures of the BCS pairing in light-stimulated 
transitions of photons between two intersecting nonparallel
laser beams. We demonstrate the dramatic dependence of the Bragg
diffraction rate on the BCS order parameter. 
The Bragg diffraction of atoms has already been experimentally
used as a method to split a BE condensate \cite{KOZ99} and to 
perform spectroscopic measurements on the condensates \cite{STE99,STA99}.

In this paper we theoretically study the optical response 
of a superfluid state in a zero-temperature FD gas.
First, we consider the propagation of low-intensity light and
calculate the leading low-density correction to the standard
century-old column density result of the electric susceptibility,
also known as the Clausius-Mossotti relation \cite{JAC75}.
This correction is a direct consequence of the quantum-statistical 
position correlations between different atoms that modify the
optical interactions at small interatomic separations. 

FD statistics exhibits a short-range ordering of atoms in the gas.
In the absence of a superfluid state the FD statistics forces a
regular spacing between the atoms in the same internal state 
within the characteristic correlation length 
$\xi_{\up\up}\sim 1/k_F$. Here $k_F$ denotes the Fermi wave number.
Fermionic atoms repel each other
and short interatomic separations are suppressed by the Pauli
exclusion principle. The pair correlation function between two
atoms displays antibunching.

The resonant dipole-dipole interactions between different atoms,
which behave as $1/r^3$, are dominant at small interatomic separations.
The Fermi repulsion suppresses these optical interactions and 
therefore the incoherent scattering of light in the atomic sample
is reduced. As a result a FD
gas exhibits a dramatic narrowing of the absorption line for the 
coherently scattered light \cite{JAV99,RUO99c}.

In the $s$-wave BCS transition the particles near the Fermi surface 
having opposite momenta and different internal quantum numbers 
tend to appear in pairs. This leads, e.g., to a finite energy
gap $\Delta$ in the excitation spectrum of the system and to a 
nonvanishing expectation value of the anomalous correlation function
for the matter-field annihilation operators $\< \psi_\uparrow (\rv)
\psi_\downarrow (\rv) \>$. Here the two internal states are referred to
as $\up$ and $\down$.

The BCS $s$-wave pairing 
also introduces short-range ordering of atoms within the characteristic
correlation length $\xi_{\up\down}\sim \epsilon_F/(\Delta k_F)$ 
between atoms in different internal states \cite{LIF80}.
Here $\epsilon_F$ denotes the Fermi
energy. Due to the macroscopic two-particle coherence the atoms in
different internal levels attract each other and the BCS pairing enhances
small interatomic separations. The pair correlation function between two
atoms with different quantum numbers displays bunching. As a result
the optical interactions and the incoherent scattering of light in the
atomic gas are {\it enhanced}. This broadens the absorption line for 
the coherently scattered light.

As a second topic we study the Bragg spectroscopy of a FD gas with
BCS pairing. In the Bragg spectroscopy an optical potential couples
to the local number density of atoms and connects two external atomic 
states of different momenta.
The scattering rate of atoms is proportional
to the dynamical structure function of the atomic gas 
\cite{JAV95b,STE99,STA99}.
The structure function has been extensively studied for the
case of incoherently scattered light in optically thin BE 
\cite{JAV95b,GRA96,RUO97b,CSO98} and FD \cite{JAV95b,MAR98,BUS98} atomic gases.

The structure function contains distinct quantum-statistical
features in the case of fermionic and bosonic atomic gases \cite{JAV95b}.
For a FD gas it demonstrates the Fermi {\it inhibition} of the 
spontaneous scattering of photons \cite{JAV95b,MAR98,BUS98}.
The structure function may also be used as a method of determining 
the relative phase between two BE condensates \cite{RUO97b,PIT99}.

In this paper we show that a superfluid state may significantly increase
the value of the static structure function. This is because due to
the BCS pairing atoms and holes near the Fermi surface are mixed;
with a given recoil momentum there exist more unoccupied states
to which atoms can scatter. The effect of superfluidity is stronger 
for nearforward scattering.

In Sec.~\ref{sec:ground} we introduce the Hamiltonian density for
ground-state atoms in the absence of the driving light field. The analysis
of the quasiparticles follows the standard BCS theory \cite{LIF80}.
We emphasize the effect of the quantum statistics on the pair correlation
function. In Sec.~\ref{sec:prop} we study the propagation of light
in a FD atomic gas. The interaction between light and matter is discussed
in general terms in Sec.~\ref{sec:bas}. The atomic polarization is solved for
the low-density atomic gas, complete with the dependence on the atomic
level scheme and on various light polarizations in Sec.~\ref{sec:pola}.
The electric susceptibility representing the damping and the phase velocity
of the light beam is obtained in Sec.~\ref{sec:sus}.
The effects of the quantum statistics and the $s$-wave interactions
manifest themselves in two distinguishable parameters. A few 
remarks about the light propagation are made in Sec.~\ref{sec:sum}.
In Sec.~\ref{sec:bragg} we consider the possibilities to observe the
BCS pairing via Bragg spectroscopy. The Bragg diffraction of atoms
in the laser field probes the structure function of the gas.
In Sec.~\ref{sec:stru} we calculate the structure function for a superfluid
state. A few concluding remarks are made in Sec.~\ref{sec:con}.

\section{Ground-state atoms}
\label{sec:ground}

\subsection{Hamiltonian}
\label{sec:ham}

We assume a FD gas occupying two different internal sublevels
$|g,\up\>$ and $|g,\down\>$ of the same atom with electronically excited
levels $|e,\nu\>$.
In the absence of the driving light field, atoms in the electronic ground
state are described in second quantization
by the Hamiltonian density $\Hc_g$ \cite{LIF80,FET71}:
\beq
\Hc_g=\sum_\nu \psi^\dagger_{g\nu}(H^{g\nu}_{\rm c.m.}-\mu_{g\nu})
\psi_{g\nu}+ \hbar u_g \psi^\dagger_{g\up}\psi^\dagger_{g\down}
\psi_{g\down}\psi_{g\up}\,,
\label{ground}
\eeq
where $\psi_{g\nu}(\rv t)$ denotes the atom-field annihilation operator for 
level $|g,\nu\>$
in the Heisenberg picture, $\mu_{g\nu}$ is the corresponding chemical
potential, and $H^{g\nu}_{\rm c.m.}$ stands for the center-of-mass (c.m.)
Hamiltonian.
We have approximated the finite-range interparticle potential by
a contact interaction with the strength given by 
$u_g=4\pi a_g\hbar/m$. Here $a_g$ and $m$
denote the $s$-wave scattering length and
the mass of the atom. The atoms in different internal states
can interact via $s$-wave scattering. On the other hand, due to the
Pauli exclusion principle
there only is a very weak $p$-wave scattering between two atoms in the same
level, which is ignored in Eq.~{(\ref{ground})}.

\subsection{BCS pairing}
\label{sec:bcs}

Before the light is switched on, the system is described by the
Hamiltonian density $\Hc=\Hc_g$ [Eq.~{(\ref{ground})}]. 
The assumption that the driving light only weakly disturbs the 
system allows us to evaluate the electric susceptibility by using
the ground-state atom correlations determined by 
$\Hc_g$, even in the presence of the driving light. 
We assume a homogeneous atomic sample and introduce a
plane-wave basis for the field operators: 
\beq
\psi_{g\nu}(\rv)={1\over \sqrt{V}}
\sum_{\kv}b_{\kv\nu}e^{i\kv\cdot \rv}\,.
\eeq 
In the Hamiltonian {(\ref{ground})} we introduce the standard canonical 
transformation to the Bogoliubov quasiparticles \cite{LIF80,FET71}  
\begin{mathletters}
\bea
\alpha_\kv &=& u_\kv b_{\kv\down}-v_\kv b_{-\kv\up}^\dagger ,\\
\beta_{-\kv} &=& u_\kv b_{-\kv\up}+v_\kv b_{\kv\down}^\dagger\,,
\eea
\label{bogo}
\end{mathletters}
where $u_\kv$ and $v_\kv$ are real, depend only only on $|\kv|$, and satisfy
$u_\kv^2+v_\kv^2=1$. 
The requirement that linearized mean field fluctuations of $\Hc_g$ 
in Eq.~{(\ref{ground})} be diagonal in the
quasiparticle representation sets an additional constraint and we obtain
\beq
u_\kv^2 = {1\over 2}\(1+{\xi_\kv\over E_\kv}\), \quad
v_\kv^2 = {1\over 2}\(1-{\xi_\kv\over E_\kv}\) \,,
\label{bogo2}
\eeq
where $E_\kv=\sqrt{\Delta^2+\xi_\kv^2}$, $\xi_\kv =\epsilon_{\kv}-
\bar{\mu} +\hbar u_g (\rho_\up+\rho_\down)/2$, and the energy gap
\beq 
\Delta=-{\hbar u_g\over V} \sum_\kv u_\kv v_\kv(1-\bar{n}^q_{\alpha\kv}
-\bar{n}^q_{\beta\kv})\,.
\eeq 
In equilibrium, the quasiparticle occupation numbers 
$\bar{n}^q_{\alpha\kv}\equiv \< \alpha_\kv^\dagger \alpha_\kv \>$ and
$\bar{n}^q_{\beta\kv}\equiv \< \beta_\kv^\dagger \beta_\kv \>$ satisfy
FD statistics with $\bar{n}^q_{\alpha\kv}=\bar{n}^q_{\beta\kv}=
(e^{E_\kv/k_BT}+1)^{-1}$.
The dispersion relation for free particles
is given by $\epsilon_{\kv}=\hbar^2 k^2/(2m)$ and
the average of the chemical potentials is $\bar{\mu}=(\mu_\up+\mu_\down)/2$.
For simplicity, we assume $\mu_\up=\mu_\down$.
For the gap parameter at $T=0$ we use the weak-coupling
approximation $\Delta\simeq 1.76 k_B T_c$ \cite{LIF80,HOU97}, where
\beq
k_BT_c\simeq {8\epsilon_F\over\pi} e^{\gamma-2}
\exp\[ -{\pi\over2k_F|a_g|}\] \,.
\eeq
Here the Fermi wave number $k_F=(6\pi^2\rho)^{1/3}$ is defined in terms of
the atom density $\rho$, $T_c$ denotes the critical temperature of the BCS
phase transition, and  $\gamma\simeq0.5772$.

\begin{figure}
\begin{center}
\epsfig{file=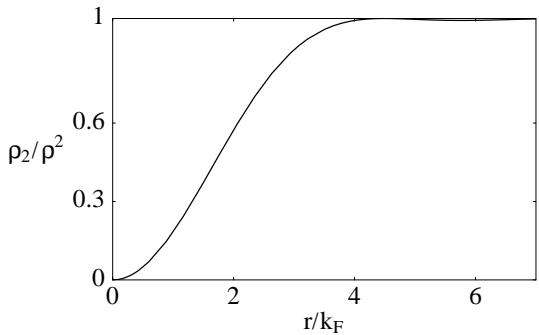,width=7.5cm}
\end{center}
\caption{
The pair correlation function for an ideal homogeneous Fermi-Dirac gas
at zero temperature. The two atoms have the same internal quantum numbers.
The pair correlation function represents the joint probability distribution
for the position of the second atom given that the first atom sits at
the origin. Two fermionic atoms repel each other and small interatomic
separations are suppressed.
}
\label{fig:pair}
\end{figure}

In the superfluid phase transition the atoms in the different
internal states $\up$ and $\down$ form quasiparticle pairs resulting
in a nonvanishing anomalous correlation 
$\<\psi_\up(\rv_1)\psi_\down(\rv_2)\>$.
The effect of this macroscopic two-particle coherence on the atomic
position correlations is particularly transparent in the case of the pair 
correlation function:
\beq
\rho_2(\rv_1\nu\eta,\rv_2\sigma\tau)\equiv \< \psi^\dagger_{g\nu}(\rv_1)
\psi^\dagger_{g\sigma}(\rv_2)\psi_{g\tau}(\rv_2)
\psi_{g\eta}(\rv_1) \>\,.
\label{pair}
\eeq
In the ground state of $\Hc_g$ [Eq.~{(\ref{ground})}],
determined by the vacuum of 
the Bogoliubov quasiparticles [Eq.~{(\ref{bogo})}],
the pair correlation function reads (for $\nu\neq\sigma$)
\begin{mathletters}
\beq
\rho_2(\rv_1\nu\nu,\rv_2\sigma\sigma) = \rho_\nu\rho_\sigma+
|\< \psi_{g\nu}(\rv_1)\psi_{ g\sigma}(\rv_2) \>|^2,\label{bcor}
\eeq
\beq
\rho_2(\rv_1\nu\nu,\rv_2\nu\nu) = \rho_\nu^2-
|\< \psi^\dagger_{g\nu}(\rv_1)\psi_{ g\nu}(\rv_2) \>|^2\,.\label{acor}
\eeq
\label{gaus}
\end{mathletters}
The pair correlation function represents the joint probability 
distribution for the positions of two atoms. We note from
Eq.~{(\ref{acor})} that the fermionic atoms in the same internal state
repel each other analogously to antibunching of photons and short 
interatomic separations are inhibited. For an ideal homogeneous
FD gas, in the absence of a superfluid state, we can analytically 
evaluate Eq.~{(\ref{acor})}. We obtain \cite{RUO99c}
\beq
\rho_2(r;\nu\nu,\nu\nu) = \rho^2\,\left\{ 1-{9\over k_F^4 r^4}\[ 
{\sin k_Fr\over k_Fr}-\cos k_Fr\]^2\right\} \,,
\label{eq:DCF}
\eeq
where $r\equiv |{\bf r}_1-{\bf r}_2|$. In Fig.~\ref{fig:pair} 
we have shown the pair
correlation function (\ref{eq:DCF}). The FD repulsion between different
atoms results in suppressed dipole-dipole interactions and in the Fermi
{\it inhibition} of incoherently scattered light~\cite{JAV99,RUO99c}.

As a result of the BCS pairing atoms in different internal
states attract each other analogously to bunching of photons according 
to Eq.~{(\ref{bcor})}. Therefore short interatomic separations are enhanced.
In the next section we find that this may lead to {\it enhanced} optical
interactions and incoherent scattering of light.

\section{Light propagation}
\label{sec:prop}

\subsection{Basic relations}
\label{sec:bas}

In this section we introduce terms in the Hamiltonian that
result from the electromagnetic fields. We briefly recapitulate
and generalize our previous quantum field-theoretical analysis of the
light-matter interactions \cite{RUO97a,RUO97c,JAV99,RUO99c,RUO99d}.
As a basic assumption, atoms are represented as
point dipoles and the radiated field in a medium has the familiar 
expression of the dipolar field [Eq.~{(\ref{eq:GDF})}].

\subsubsection{Electromagnetic field}

We consider the nonrelativistic Hamiltonian formalism of electrodynamics.
It is advantagous to study the propagation of light by introducing
the dipole approximation for atoms and the corresponding 
Hamiltonian in the {\it length} gauge obtained in the Power-Zienau-Woolley 
transformation \cite{POW64,COH89,LEW94}.

The electromagnetic field introduces additional terms in the system 
Hamiltonian. The Hamiltonian for the free electromagnetic field-energy is
\beq
H_F=\int d^3r\, \Hc_F = \sum_q \hbar \omega_q a^\dagger_q a_q\,.
\eeq
Here $\omega_q$ and $a_q$ denote the mode frequency and the photon 
annihilation operator. The mode index $q$ incorporates both the wave 
vector ${\bf q}$ and the transverse polarization $\pol_q$.
In the length gauge the basic dynamical degree of freedom for the light 
field is the electric displacement $\Dv(\rv)$ which interacts with the 
atomic polarization $\Pv(\rv)$,
\beq
\Hc_D=-{1\over\eo}\,\Pv(\rv)\cdot\Dv(\rv) \,.
\label{h2}
\eeq
In the dipole approximation the polarization is given in terms of 
the density of atomic dipoles $\Pv(\rv)=\sum_i \dv_i \delta(\rv_i-\rv)$.
Here $\dv_i$ and $\rv_i$ denote the dipole operator and the c.m. 
position operator for the $i$th atom. In second quantization 
the positive frequency component of the polarization is given by
\beq
\Pv^+(\rv) = \sum_{\nu,\eta} \dv_{g\nu e\eta}\psi^\dagger_{g\nu}(\rv)
\psi_{ e\eta}(\rv) \equiv \sum_{\nu,\eta}
\Pv^+_{\nu\eta}(\rv)\,,
\label{pol}
\eeq
\beq
\Pv^+_{\nu\eta}(\rv) \equiv \dv_{g\nu e\eta}\psi^\dagger_{g\nu}(\rv)
\psi_{ e\eta}(\rv)\,,
\label{polcomp}
\eeq
where $\dv_{g\nu e\eta}$ stands for the dipole matrix element 
for the transition $|e,\eta\>\rightarrow |g,\nu\>$
\beq
\dv_{g\nu e\eta}\equiv \Dc \sum_\sigma \pol_{\sigma}
\< e\eta;1g|1\sigma;g\nu\>\,.
\label{dipole}
\eeq
Here ${\cal D}$ denotes the reduced dipole matrix element, which
is chosen to be real, and $\dv_{e\eta g\nu}=\dv_{g\nu e\eta}^*$. 
The summation in Eq.~{(\ref{dipole})} runs over the unit circular 
polarization vectors $\sigma=\pm1,0$ weighted by the 
Clebsch-Gordan coefficients of the corresponding optical transitions.
Here light fields with the polarizations $\pol_\pm$ and $\pol_0$ drive 
the transitions $|g,\nu\>\rightarrow|e,\nu\pm1\>$ and $|g,\nu\>
\rightarrow|e,\nu\>$, respectively.

For a weak external magnetic field the nuclear spin $I$, the electron
spin $S$, and the orbital angular momentum $L$ may be coupled. In
that case $F$ and $m_F$, defined by ${\bf F}={\bf I}+{\bf S}
+{\bf L}$, are good quantum numbers. With stronger magnetic fields the nuclear
spin decouples and the optical transitions are between different 
sublevels of $J$, where ${\bf J}={\bf S}+{\bf L}$. In the latter
case we may assume that the atoms occupy a single manifold of $m_I$.
In the following calculation we consider atomic sublevels determined by 
$m_F$.

The positive frequency component of the
electric field $\Ev^+$ may be expressed in terms of the positive
frequency component of the driving electric displacement ${\bf D}^+_F$, 
with the wave number $k$, and the dipole radiation field~\cite{RUO97a}: 
\begin{mathletters}
\bea
\eo{\bf E}^+({\bf r})& =& {\bf D}^+_F({\bf r}) +
\int d^3r'\,
{\sf G}({\bf r}-{\bf r'})\,{\bf P}^+({\bf r}')\,,
\label{eq:MonoD}\\
{\sf G}_{ij}({\bf r})& =& 
\left[ {\partial\over\partial r_i}{\partial\over\partial r_j} -
\delta_{ij} {\bbox \nabla}^2\right] {e^{ikr}\over4\pi r}
-\delta_{ij}\delta({\bf r})\,.
\label{eq:GDF}
\eea
\label{eq:atomlight}
\end{mathletters}
We also have the familiar relation
\beq
\Dv(\rv)=\eo \Ev(\rv)+\Pv(\rv)\,.
\eeq

Equation (\ref{eq:atomlight}) is the integral representation for the 
Maxwell's wave equation of the electric field in the presence of 
dipole atoms and the monochromatic dipole radiation kernel 
${\sf G}(\rv)$ coincides with the corresponding classical 
expression \cite{JAC75}. The explicit form of the radiated field from
a dipole with the amplitude $\hat{\bf p}$ reads:
\bea
\lefteqn{{\sf G}({\bf r})\,\hat{\bf p}=
{k^3\over4\pi}
\left\{ (\hat{\bf n}\!\times\!\hat{\bf p}
)\!\times\!\hat{\bf n}{e^{ikr}\over kr}\right.}\nonumber\\
&&\mbox{} \left. +[3\hat{\bf n}(\hat{\bf n}\cdot\hat{\bf p})-\hat{\bf p}]
\bigl[ {1\over (kr)^3} - {i\over (kr)^2}\bigr]e^{ikr}
\right\}-{\hat{\bf p}\,\delta({\bf r})\over3}\,,
\label{eq:DOL}
\eea
where $\hat{\bf n} = {{\bf r}/ r}$. The volume integral over $1/r^3$ 
in Eq.~{(\ref{eq:DOL})} is not absolutely convergent in the neighborhood 
of the origin. The expression {(\ref{eq:DOL})} should be understood 
in such a way that the integral of the term inside the curly brackets over 
an infinitesimal volume enclosing the origin vanishes \cite{RUO97a}.
We note that Eq.~(\ref{eq:MonoD}), with the correct delta function 
contribution
from Eq.~{(\ref{eq:DOL})}, yields ${\bbox \nabla}\cdot \Dv(\rv)=0$ for
neutral atoms justifying the use of the electric displacement,
instead of the electric field, in Eq.~{(\ref{h2})}.

The nonrelativistic propagator (\ref{eq:DOL}) involves an explicit
high-frequency cutoff \cite{RUO97a}. In situations where
integral expressions containing the propagator ${\sf G}(\rv)$ 
are not absolutely convergent the integrals are defined in such
a way that the integration over spherical angles should be performed
first \cite{RUO97a}.

In the length gauge the Hamiltonian also contains the 
polarization self-energy term:
\beq
\Hc_{\rm P}={1\over2\eo}\,\Pv(\rv)\cdot\Pv(\rv)\,.
\eeq
However, this is proportional to the overlap of the atomic dipoles 
and, in the limit of low light-intensity, all contact interaction terms
between different dipole atoms were shown in Ref.~\cite{RUO97c}
to be inconsequential for light-matter dynamics. Although the result is
valid for an arbitrary number of atoms, the cancelation of 
the contact interactions is especially transparent in the case of 
a superradiant decay of only two point dipoles~\cite{RUO97c}. 

\subsubsection{$S$-wave interactions}

Two cold fermionic atoms in different internal states interact by means of
$s$-wave scattering. This introduces interaction terms between
ground-state atoms in Eq.~(\ref{ground}). In the presence of driving
light we also have $s$-wave scattering between different
electronically excited levels and between electronically excited
and ground-state atoms.

We write the contribution to the Hamiltonian density that consists
of the excited-level operators as 
\bea
\Hc_{e} &=& \sum_\nu \psi^\dagger_{e\nu}(H^{e\nu}_{\rm c.m.}+
\hbar\omega_0-\mu_{e\nu})\psi_{e\nu}\nonumber\\
&&\mbox{}+{\hbar\over2} \sum_{\rm \nu\eta\sigma\tau}
u_{e\nu\eta,e\sigma\tau} \psi^\dagger_{e\nu}\psi^\dagger_{e\eta}
\psi_{e\sigma}\psi_{e\tau}\,.
\label{hee}
\eea
Here $u_{e\nu\eta,e\sigma\tau}=4\pi\hbar a_{e\nu\eta,e\sigma\tau}/m$
describes the two-body $s$-wave scattering between the atoms. For simplicity,
the frequency of the optical transition $\omega_0$ is assumed to
be independent of the atomic sublevel. For typical values of the
optical linewidth the c.m.\ motion for the excited atoms may be omitted
\cite{JAV95b}. In this paper we consider a situation where the intensity
of the driving light field is low and therefore the density of the excited
atoms is low. Hence, to leading order in the limit of low light-intensity
the second term in Eq.~{(\ref{hee})} makes no contribution to the optical 
response, and we do not address its explicit form in more detail.

We assume that to leading order all remaining interactions 
between the ground-state and excited-state atoms, which cannot be 
accounted for when the atoms are modeled as point dipoles,
are governed by the following interactions \cite{LEW94}:
\beq
\Hc_{ge} = \hbar
\sum_{\nu\eta\sigma\tau}u_{g\nu\eta e\sigma\tau}\psi^\dagger_{g\nu}
\psi^\dagger_{e\eta}\psi_{e\sigma}\psi_{g\tau}\,.
\label{hge}
\eeq
If we assume the conservation of the angular momentum of the colliding
particles, the two-body interaction can be written in the following
form \cite{HO98}:
\beq
\Hc_{ge} = \hbar \sum_{F=|f_a-f_b|}^{f_a+f_b}
\sum_{m_F=-F}^{F} u_{F,m_F}{\cal O}_{Fm_F}^\dagger{\cal O}_{Fm_F}\,,
\label{ho}
\eeq
where $u_{F,m_F}=4\pi\hbar a_{F,m_F}/m$ and
\beq
{\cal O}_{Fm_F}\equiv \sum_{m_am_b} \<Fm_F;f_af_b|f_a,m_a;f_b,m_b\>\,
\psi_{f_am_a}\psi_{f_bm_b}\,.
\eeq
Here $ \<Fm_F;f_af_b|f_a,m_a;f_b,m_b\>$ is the Clebsch-Gordan coefficient
for forming the state $|Fm_F\>$ from two colliding particles in
the angular momentum states $|f_a m_a\>$ and $|f_bm_b\>$.

As an example we consider the $f=1/2\rightarrow 3/2$ transition.
By expanding Eq.~{(\ref{ho})} we obtain $\Hc_{ge}=\Hc^A_{ge}+
\Hc^B_{ge}$, where $\Hc^A_{ge}$ does not mix the population between
the different sublevels. The explicit form reads
\beq
\Hc^A_{ge} = \hbar\sum_{\nu,\sigma}
u_{g\nu,e\sigma}\psi_{e\sigma}^\dagger \psi_{g\nu}^\dagger
\psi_{g\nu}\psi_{e\sigma}\,.
\label{hega}
\eeq
For the $f=1/2\rightarrow 3/2$ transition the summation runs over
$\nu=\pm1/2$ and $\sigma=\pm1/2,\pm3/2$.
The interaction strengths $u_{g\nu,e\sigma}$ are determined
in terms of the interaction strenghts $u_{F,m_F}$ for the scattering
channels $|Fm_F\>$:
\bea
u_{g\pm {1\over2},e\pm{3\over2}} &=& u_{2,\pm2}, \nonumber\\
u_{g\pm {1\over2},e\mp {1\over2}} &=& {1\over2} (u_{1,0}+u_{2,0}),\nonumber\\
u_{g\pm {1\over2},e\mp{3\over2} } &=& {3\over4} u_{1,\mp 1}+{1\over4} 
u_{2,\mp 1},\nonumber\\
u_{g\pm {1\over2},e\pm{1\over2} } &=& {1\over4} u_{1,\pm 1}+
{3\over4} u_{2,\pm 1}.\nonumber
\eea
On the other hand, $\Hc^B_{ge}$ consists of spin-exhange collision terms:
\bea
&&\Hc^B_{ge}/\hbar = {1\over2}(u_{2,0}-u_{1,0})\,
\psi_{e{1\over2}}^\dagger \psi_{g-{1\over2}}^\dagger\psi_{g{1\over2}}
\psi_{e-{1\over2}} \nonumber\\&&
\mbox{}+{\sqrt{3}\over4}(u_{2,1}-u_{1,1})\,
\psi_{e{3\over2}}^\dagger \psi_{g-{1\over2}}^\dagger
\psi_{g{1\over2}}\psi_{e{1\over2}}
\nonumber\\&&
\mbox{}+{\sqrt{3}\over4}(u_{2,-1}-u_{1,-1})\,
\psi_{e-{3\over2}}^\dagger \psi_{g{1\over2}}^\dagger
\psi_{g-{1\over2}}\psi_{e-{1\over2}}+ {\rm H.c.} \,.
\eea
Similar nonlinear wave-mixing terms can have interesting effects
on the dynamics of spinor BE condensates \cite{LAW98}.

If the $s$-wave scattering amplitude $u_{F,m_F}$ is independent of
$F$ we obtain $\Hc^B_{ge}=0$. Furthermore, if the scattering 
is also independent of $m_F$,  $u_{g\nu,e\sigma}$ in Eq.~{(\ref{hega})}
does not depend on sublevels $\nu$ and $\sigma$.
For simplicity, in the following we assume that $\Hc^B_{ge}=0$.

\subsection{Atomic polarization}
\label{sec:pola}

The dipole radiation [Eq.~{(\ref{eq:atomlight})}] describes the scattered
light in a medium. In this section we study the optical response 
of an atomic gas by solving nonperturbatively the polarization of 
the matter-field for 
a low-density atomic gas in the limit of low light-intensity. The main
items are the steady-state equation of the atomic polarization
[Eq.~{(\ref{p1})}] and its solution by means of the 
low-density decorrelation approximation [Eq.~{(\ref{p2})}].

\subsubsection{Equation of motion}

We consider the optical response of the atomic polarization in the 
limit of low light-intensity. We derive the corresponding Heisenberg 
equations of motion for the matter-field operators. Alternative 
approaches have used, e.g., the Schwinger-Keldysh techniques \cite{FLE99}.
We assume that the spin-exchanging collisions between the ground-state
and the excited-state atoms vanish indicating $\Hc_{ge}=\Hc_{ge}^A$.
Then the Heisenberg equations of motion for atomic field operators
are obtained from the Hamiltonian densities $\Hc_g$ [Eq.~{(\ref{ground})}],
$\Hc_D$ [Eq.~{(\ref{h2})}], $\Hc_e$ [Eq.~{(\ref{hee})}], and
$\Hc^A_{ge}$ [Eq.~{(\ref{hega})}].

In the limit of low light-intensity we derived from the Heisenberg
equations of motion a hierarchy of equations for correlation functions 
involving atomic polarization and atom density \cite{RUO97a,RUO97c}. 
In the case of the present system we may proceed similarly. 
As far as the optical response is concerned it is again assumed 
that we can concentrate on the dynamics of internal degrees of freedom 
for the atoms and the light. Hence, in the equation of motion for 
the atomic polarization the kinetic energy of the atoms is neglected. 
Nevertheless, the quantum statistics of the different c.m. motional 
states is still fully included.

The vacuum electromagnetic fields are eliminated by performing the
field-theory version of the Born and Markov approximations~\cite{RUO97a}.
We then obtain the equation of motion for the polarization operator 
component $\Pv^+_{\nu\eta}(\rv)$,
\begin{eqnarray}
\lefteqn{{d\over dt}\,\Pv^+_{\nu\eta} =
(i\delta-\gamma)\Pv^+_{\nu\eta}
+ i\kappa \psi^\dagger_{g\eta}\psi_{g\sigma} {\sf P}^{\nu\eta}_{\eta\sigma}
\cdot {\bf D}^+_F} \nonumber\\
&&\mbox{}+ i\kappa
\int d^3r'\,{\sf P}^{\nu\eta}_{\eta\sigma}\cdot
{\sf G}'({\bf r}-{\bf r'})\psi^\dagger_{g\nu}\Pv^+(\rv')
\psi_{g\sigma}\nonumber\\
&&\mbox{} -i\kappa
\int d^3r'\,{\sf P}^{\nu\eta}_{\sigma\nu}\cdot
{\sf G}'({\bf r}-{\bf r'})\psi^\dagger_{e\sigma}\Pv^+(\rv')
\psi_{e\nu}\nonumber\\
&& \mbox{}- i\kappa 
\psi^\dagger_{e\sigma}\psi_{e\nu}{\sf P}^{\nu\eta}_{\sigma\nu}\cdot 
{\bf D}^+_F +iu_g (1-\delta_{\nu,\sigma})\, \psi^\dagger_{g\nu}
\psi^\dagger_{g\sigma}\psi_{g\sigma}\psi_{e\eta}\nonumber\\
&& \mbox{}+ iu_{g\nu,e\sigma} \psi^\dagger_{g\nu}
\psi^\dagger_{e\sigma}\psi_{e\sigma}\psi_{e\eta}-
iu_{g\sigma,e\eta} \psi^\dagger_{g\nu}
\psi^\dagger_{g\sigma}\psi_{g\sigma}\psi_{e\eta}
\nonumber\\ && \mbox{}+ i(u_{e\xi\eta,e\sigma\tau}-
u_{e\eta\xi,e\sigma\tau}) \psi^\dagger_{g\nu}
\psi^\dagger_{e\xi}\psi_{e\sigma}\psi_{e\tau}\,.
\label{eq:CR1}
\end{eqnarray}
Here we use the notational convention that the repeated indices 
$\sigma$, $\tau$, and $\xi$ indicate the summation over the corresponding
sublevels.
In Eq.~{(\ref{eq:CR1})} we have shown explicitly only the nonlocal 
position dependence.
The atom-light detuning is denoted by $\delta$ and the Kronecker delta
function by $\delta_{\nu,\sigma}$. The spontaneous 
linewidth $\gamma$ is given by
\beq
\gamma \equiv {\Dc^2 k^3\over 6\pi\hbar\eo}\,.
\eeq
We have also defined $\kappa\equiv \Dc^2/(\hbar\eo)$ in terms of
the reduced dipole matrix element ${\cal D}$.
In Eq.~{(\ref{eq:CR1})} we introduced the following projection operator 
\bea
\lefteqn{{\sf P}^{\nu\eta}_{\xi\tau} \equiv {\dv_{g\nu e\eta}
\dv_{e\xi g\tau}\over {\cal D}^2}}\nonumber\\
&&= \sum_{\sigma_1,\sigma_2}
\pol_{\sigma_1}\pol_{\sigma_2}^* \<e\eta;1g|1\sigma_1;g\nu\>
\<e\xi;1g|1\sigma_2;g\tau\>\,,
\label{pro}
\eea
to include the dependence of the scattered light on the polarizations 
and on the atomic level structure. 

In the present formalism the atoms are represented by ideal point dipoles
which may overlap. Obviously, real atoms with a hard-core interatomic 
potential cannot overlap.
We remove the contact dipole-dipole interactions between different
atoms. This is done by introducing in Eq.~{(\ref{eq:CR1})} the propagator 
\beq
{\sf G}_{ij}'(\rv)\equiv {\sf G}_{ij}(\rv)+
{\delta_{ij}\delta(\rv)\over3}\,.
\label{subs}
\eeq
that explicitly cancels the contact interaction of ${\sf G}(\rv)$ 
displayed in Eq.~{(\ref{eq:DOL})}.
The purpose of this definition is to yield a vanishing integral 
for ${\sf G}'(\rv)$ over an infinitesimal volume enclosing the 
origin. As shown for the case of low-intensity light in 
Ref.~\cite{RUO97c}, we can make the substitution 
(\ref{subs}) without changing the outcome of the optical response,
and without any additional assumption of the hard-core interatomic
potentials, even for ideal point dipoles.
This is because due to the divergent dipole-dipole
interactions all correlation functions for atomic dipoles vanish whenever 
two position arguments are the same. As a result the contact interaction
terms between different dipole atoms do not have any effect on the
light-matter dynamics. The independence of the optical
response of the collection of dipole atoms
on the substitution (\ref{subs}) is the underlying explanation
for the Lorentz-Lorenz local-field correction to the electric 
susceptibility.

We consider the limit of low intensity of the driving light.
Obviously, light has to be present in order to produce population in the 
electronically excited levels and an excited-state field amplitude is
proportional to the light field amplitude. Therefore, we take the 
low-intensity limit by retaining only those products of operators in 
Eq.~{(\ref{eq:CR1})} that involve at most one excited state field 
operator or the electric displacement amplitude \cite{RUO97a}. 
Then, e.g., the third line and the last line in Eq.~{(\ref{eq:CR1})} 
make no contribution to the equation of motion for $\Pv^+(\rv)$
for low light-intensity.

We also note that in the low-intensity limit the pair correlation
function (\ref{pair}) is determined by the Hamiltonian density
in the absence of the driving light $\Hc_g$ [Eq.~{(\ref{ground})}].
This is because the effect of the light on the ground-state field
amplitudes involves terms that contain at least one excited-state 
field operator {\it and} one light-field amplitude:
$$
\dot\psi_g (\rv)\propto \dv_{ge}\cdot\Dv^-(\rv)\psi_e(\rv)\,.
$$
Thus, to leading order in the low-intensity limit the effect
of the driving light on the ground-state atom correlation functions
vanishes.

For the expectation value of the polarization we use the notation 
$\Pv_{1\nu\eta}\equiv \<\Pv^+_{\nu\eta}\>$, with $\nu$ and $\eta$ 
denoting the atomic sublevel. 
The steady-state solution of $\Pv_{1\nu\eta}$ in the limit of low
light-intensity is given by
\bea
\lefteqn{
\Pv_{1\nu\eta}(\rv_1)=\alpha\rho_{\nu }{\sf P}^{\nu\eta}_{\eta\nu}
\cdot\Dv^+_F(\rv_1)+
\sum_\sigma\Fc^{\sigma\eta}_{\sigma\nu}
\Pv_2(\rv_1\sigma\sigma;\rv_1\nu\eta )}
\nonumber\\ &&+ \mbox{}
\alpha\sum_{\sigma\tau\xi}
\int d^3 r_2 \, {\sf P}^{\nu\eta}_{\eta\sigma}\cdot{\sf G'}(\rv_1-\rv_2)
\Pv_2(\rv_1\nu\sigma;\rv_2\tau\xi )\,.
\label{p1}
\eea
Here $\alpha=-\Dc^2/[\hbar\eo(\delta+i\gamma)]$ is the 
polarizability of an isolated atom and
$\rho_\nu\equiv \<\psi^\dagger_{g\nu}\psi_{g\nu}\>$ denotes 
the ground-state atom density in level $\nu$. We have also defined
\beq
\Pv_2(\rv_1\nu\eta;\rv_2\sigma\tau) \equiv \< \psi^\dagger_{g\nu}(\rv_1)
\Pv^+_{\sigma\tau}(\rv_2)\psi_{g\eta}(\rv_1)\>\,,
\eeq
\beq
\Fc^{\sigma\eta}_{\tau\nu} \equiv {1\over \delta+i\gamma}[u_{g\sigma e\eta}
-(1-\delta_{\tau\nu})u_g]\,.
\eeq
The normally ordered expectation value $\Pv_2(\rv_1\nu\eta;\rv_2\sigma\tau)$ 
describes correlations between an atomic dipole at $\rv_2$ and a 
ground-state atom at $\rv_1$. In the integral of Eq.~{(\ref{p1})} it
represents a process in which an excited atom at $\rv_2$ emits a photon 
and excites a ground-state atom at $\rv_1$.
The tensor $\Fc^{\sigma\eta}_{\tau\nu}$ 
generates the collisionally-induced level shifts.

\subsubsection{Low-density approximation}

So far, we have obtained a steady-state solution for the atomic 
polarization {(\ref{p1})} that acts as a source for the secondary
radiation in Eq.~{(\ref{eq:MonoD})}. Equation {(\ref{p1})} involves
unknown correlation function $\Pv_2$. Basically, we could continue
the derivation and obtain the equations of motion for $\Pv_2$ and
for the higher-order correlation functions. This would eventually result 
in an infinite hierarchy of equations analogous to the equations in Ref.
\cite{RUO97a}. However, even in the case of a simple level structure 
and in the absence of the $s$-wave interactions the solution for the whole 
system by stochastic simulations is demanding on computer time 
\cite{JAV99}. In the studies of the refractive index of a quantum 
degenerate BE gas Morice {\it et al.} \cite{MOR95,TIG90} 
considered a density expansion in terms of the number of atoms 
repeatedly exchanging a photon by introducing certain approximations
to the ground-state atom correlations.
Although the lowest-order density correction to the 
susceptibility of a zero-temperature FD gas may be obtained
analytically \cite{RUO99c}, in the presence of highly nontrivial quantum
statistical position correlations a rigorous density expansion is
in most cases a very challenging task. 
In this paper we consider low atom densities (in terms of $\rho/k^3$) and 
approximate 
Eq.~{(\ref{p1})} by the decoupling that is analogous to the lowest-order 
correction in Ref. \cite{MOR95},
\beq
\Pv_2(\rv_1\nu\eta;\rv_2\sigma\tau)\simeq
{\rho_2(\rv_1\nu\eta,\rv_2\sigma\sigma)\over \rho_\sigma}
\Pv_{1\sigma\tau}(\rv_2)\,,
\label{p2}
\eeq
where the ground-state pair correlation function $\rho_2$ is defined by
Eq.~{(\ref{pair})}.

The decorrelation approximation (\ref{p2}) introduces the lowest-order
correction to the optical response in terms the number of microscopic
optical interaction processes between the atoms by ignoring the
repeated scattering of a photon between the same atoms \cite{TIG90}. 
As shown in 
Ref.~\cite{RUO99c} in the absence of a superfluid state it also 
correctly generates the leading low-density correction.
The predictions of the expansion by Morice {\it et al.} \cite{MOR95} 
were tested for a zero-temperature FD gas in one dimension \cite{JAV99}. 
The agreement with the exact solution obtained by the numerical 
simulations was found to be semiquantitative and in the low-density 
limit excellent.

The dependence of the light propagation on the density fluctuations
may now be observed by inserting Eq.~{(\ref{p2})} into Eq.~{(\ref{p1})}.
If the emitting atom at $\rv_2$ and the absorbing atom at $\rv_1$ 
have the same internal state, the pair correlation function displays
repulsion and is determined by Eq.~{(\ref{acor})}. In the case of 
different internal
states the atoms attract each other and the pair correlation function
is obtained from Eq.~{(\ref{bcor})}.

It is crucial for the low-density limit that the atom operators 
in Eq.~{(\ref{eq:CR1})}
were arranged to normal order. Otherwise, commutators are generated
for higher-order correlation functions $\Pv_2,\ldots$ that 
could be of the same order in atom density as the terms in the
equation for $\Pv_{1\nu\eta}$.

\subsection{Electric susceptibility}
\label{sec:sus}

In the previous section we obtained the steady-state solution for
the atomic polarization {(\ref{p1})} by means of the low-density
approximation {(\ref{p2})}.
The optical response may now be evaluated by eliminating $\Dv_F^+$ and
$\Pv_2$ from Eqs.~{(\ref{eq:MonoD})}, {(\ref{p1})}, and {(\ref{p2})}.
As an example we calculate the vector components of $\Pv_1$ for
the $f=1/2\rightarrow 3/2$ transition having the electronically excited
sublevels $m_f=\pm1/2,\pm3/2$. The pair correlation function
in Eq.~{(\ref{p2})} is nonvanishing only with $\nu=\eta$.
Because we are dealing with a linear theory, the electric field and the
polarization are related by the susceptibility as 
$\Pv^+=\eo\chi \Ev^+$.
We consider a situation where FD gas fills the half-infinite
space $z>0$. For simplicity, we assume equal and constant atom 
densities for the spin states $\rho_\up=\rho_\down\equiv\rho$. 
To simplify further, we assume that scattering length $a_{g\nu e\sigma}$
is independent of $\nu$ and $\sigma$ corresponding to the case that the
$s$-wave interactions in Eq.~{(\ref{hega})} are independent of the
scattering channel $|Fm_F\>$. We write the incoming free field as 
a plane wave
\beq
\Dv_F(\rv)= D_F \,\pol\, e^{ikz}\,.
\eeq
We assume that it is linearly polarized with $\pol$ parallel 
to $\dv_{g{1\over2} e{1\over2}}$. By choosing $\pol\equiv\pol_x$
we have the following representation for the circular polarization vectors
in terms of the unit Cartesian coordinate vectors
\begin{equation}
\hat{\bf e}_\pm=\mp\frac{1}{\sqrt{2}}(\hat{\bf e}_y\pm i\hat{\bf e}_z),\quad
\hat{\bf e}_0 = \hat{\bf e}_x\,.
\label{eqa:UNV}
\end{equation}
With the ansatz $\Pv_{1\nu\nu}(\rv)=P\, \pol\exp{(ik'z)}$, for Im$(k')>0$,
we then immediately see that $\Pv_{1\nu\eta}=0$ for $\nu\neq\eta$.
Finally, by using Eq.~{(\ref{gaus})},
and by ignoring the effects of the surface of the atomic gas 
\cite{JAV99}, we obtain a spatially constant susceptibility for
the sample as
\begin{equation}
\chi={{k'}^2\over k^2} -1= {2{\cal C}\alpha\rho\over1-2{\cal C}
\alpha\rho/3+\Sigma_1+\Sigma_2}\,,
\label{sus}
\end{equation}
with
\bea
\Sigma_1 &=&-{{\cal C}\alpha\over\rho}\int d^3r\,e^{-ikz}\,\pol^*\cdot
{\sf G'}(\rv)\cdot\pol \[ |\< \psi_{g\up}(\rv)\psi_{ g\down}(0) \>|^2\right.
\nonumber\\
&&\left.-|\< \psi^\dagger_{g\down}(\rv)\psi_{ g\down}(0) \>|^2\],
\label{c}
\\
\Sigma_2 &=& -{1\over\rho}\sum_{\sigma} \Fc^{\up\sigma}_{\up\sigma}
\rho_2(\rv\up,\rv\sigma)\,.
\label{colshift}
\eea
Here we have used the obvious relation
$\rho_2(\rv_1\sigma,\rv_2\nu)\equiv \rho_2(\rv_1\sigma\sigma,
\rv_2\nu\nu)=\rho_2(\rv_1\nu\nu,\rv_2\sigma\sigma)$.
The parameter ${\cal C}$
denotes the value of the Clebsch-Gordan coefficient 
${\cal C}=|\<e{1\over2};1g|10;g{1\over2}\> |^2=2/3$ in the case
of the $f=1/2\rightarrow 3/2$ transition. By writing $\bar\rv\equiv k\rv$
the propagator in Eq.~{(\ref{c})} has the following expression in the
spherical coordinates
\bea
\pol_x^*\cdot {\sf G}'(\bar\rv)\cdot\pol_x &=&
{k^3 e^{i\bar{r}}\over4\pi}\left[ (1-\sin^2\theta\cos^2\phi)
{1\over\bar{r}}\right.\nonumber\\
&&\left.\mbox{}+ (3\sin^2\theta\cos^2\phi-1)({1\over
\bar{r}^3}-{i\over \bar{r}^2})\right]\,.
\label{proj}
\eea

In Eq.~{(\ref{sus})} $\Sigma_1$ is solely generated by the 
quantum-statistical position correlations between different atoms.
The effect of $s$-wave interactions is encapsulated in $\Sigma_2$.
In an uncorrelated atomic sample the atomic positions are
statistically independent and the pair correlation function
(\ref{pair}) satisfies $\rho_2(\rv\nu,\rv'\sigma)=\rho_\nu
\rho_\sigma$ resulting in $\Sigma_1=0$. For the case of uncorrelated 
atoms, and in the absence of the $s$-wave scattering,
we would obtain Eq.~{(\ref{sus})} with $\Sigma_1=\Sigma_2=0$. 
This is the standard column density
result stating that susceptibility equals polarizability of an
atom times atom density. Equation {(\ref{sus})} also involves 
the Lorentz-Lorenz local-field correction in the denominator.

The quantum-statistical corrections to the column density result
are introduced by $\Sigma_1$. It describes
the modifications of the optical interactions between neighboring
atoms due to the position correlations.
The second term in Eq.~{(\ref{c})} represents the quantum-statistical 
contribution to the scattering process in which a photon 
emitted by an atom in internal state $\nu$ at position $\rv$
is reabsorbed by another atom in internal state $\nu$ and 
located at the origin. 
According to FD statistics two fermions with the same quantum 
numbers repel each other and FD statistics forces a regular 
spacing between the atoms.
The optical interactions are dominantly generated at small 
interatomic distances and the
corrections to the susceptibility due to the second term 
in Eq.~{(\ref{c})} correspond to {\it inhibitited} multiple 
scattering of light resulting in suppressed diffusive radiation. 
In the absence of a superfluid state
FD gas exhibits a dramatic narrowing of the absorption linewidth for
coherently scattered light~\cite{JAV99,RUO99c}.

The first term in Eq.~{(\ref{c})}
represents the quantum-statistical corrections to the reabsorption
process between atoms in different internal states due to the 
two-particle coherence. This term is nonvanishing
only in the presence of a superfluid state. Because the total spin
of an interacting atom pair in Eq.~{(\ref{ground})} 
is an integer, the pairs behave as bosons \cite{LIF80}. According to
the Bose-Einstein statistics two bosons attract each other and
the BCS pairing favors small interatomic spacing. This results in
{\it enhanced} optical interactions and incoherent scattering of
light.

The electric susceptibility exhibits a Lorentzian line shape.
The optical line shift ${\cal S}$ and linewidth $\Gamma $ for 
the atomic sample are obtained from Eq.~{(\ref{sus})}
\beq
{\cal S}/\gamma = {4\pi\rho{\cal C}\over k^3}+{\cal S}_{\rm col}/\gamma
-{6\pi\over k^3}\, {\rm Re} \({\Sigma_1\over\alpha}\) \,,
\label{shift}
\eeq
\beq
\Gamma/\gamma=1-{6\pi\over k^3}\, {\rm Im}\({\Sigma_1\over\alpha}\)\,.
\label{width}
\eeq
The collisional line shift ${\cal S}_{\rm col}$, which results from 
$\Sigma_2$ [Eq.~{(\ref{colshift})}], is generated by the $s$-wave 
interactions. It depends on the BCS order parameter $\Delta$:
\bea
{\cal S}_{\rm col} &\equiv& ({u}_{g}-{u}_{ge}){\rho_2(\rv\up,\rv\down)
\over\rho}\nonumber\\
&=& \rho({u}_{g}-{u}_{ge}) \[ 1+\({\Delta\over\hbar u_g\rho}
\)^2 \] \,.
\label{colshift2}
\eea

The first term in the optical lineshift~(\ref{shift})
corresponds to the Lorentz-Lorenz local-field correction.
As far as the $s$-wave interactions can be considered local 
on the scale of the optical wavelength in Eqs.~{(\ref{ground})} 
and~(\ref{hge}) also the line shift ${\cal S}_{\rm col}$ may be
considered as a local-field shift. In that case the local-field shift 
due to the $s$-wave scattering in Eq.~{(\ref{shift})} is larger
than the Lorentz-Lorenz shift for $^6$Li if 
$$
\gamma\lesssim 210
\[1+\({\Delta\over \hbar u_g \rho}\)^2\] {(a_g-a_{ge})\over 
a_0\lambda^3}\,\,\mu m^3
s^{-1}\,.
$$
Here $\lambda$ denotes the wavelength of the incoming light and
$a_0$ is the Bohr radius.
Because $(\Delta/\hbar u_g \rho)^2$ is expected to be of the 
order of one \cite{HOU97}, the local-field shift could 
strikingly depend on the BCS order parameter $\Delta$.
The collisional line shift was recently observed for a hydrogen BE
condensate by using a two-photon $1S\rightarrow2S$ spectroscopy
\cite{KIL98}.

If the the effective range $r_u$ of the triplet $s$-wave
potential in Eqs.~{(\ref{ground})} and~(\ref{hge})
is very short, $r_u\ll 1/k$, the resonant dipole-dipole interactions
may suppress the effect of the $s$-wave scattering on the line
shift just as they cancel the effect of the polarization 
self-energy \cite{RUO97c}. However, for a metastable state,
$\gamma^{-1}$ may be large on the time scale of the atomic 
interactions. In that case the collisional shift could be
observable even for very small $r_u$.

To calculate the linewidth (\ref{width}) and line shift 
(\ref{shift}) from integral (\ref{c}) we need to evaluate 
the spatial correlation functions by using Eqs.~{(\ref{bogo})} 
and~{(\ref{bogo2})}. For instance, the expectation value for
the anomalous correlation function reads
\beq
\<\psi_\down(\rv)\psi_\up(0)\>={1\over V}\sum_\kv e^{i\kv \cdot
\rv}{\Delta\over 2E_\kv}(1-\bar{n}^q_{\alpha \kv}-
\bar{n}^q_{\beta \kv})\,.
\label{pairsum}
\eeq
The chemical potential is solved from $\rho_\nu=\rho_\nu(\bar\mu)$.
Here $\<\psi_\down(0)\psi_\up(0)\>=-\Delta/(\hbar u_g)$ is 
ultraviolet-divergent, resulting from
the assumption of the contact two-body interaction in
Eq.~{(\ref{ground})}. This interaction is momentum independent and
it is not valid at high energies. To estimate the pairing
a standard procedure is to remove the high-energy divergence 
by introducing a high-momentum cutoff~$k_c$. 
Nevertheless, we find that the optical linewidth of a FD gas
[Eq.~{(\ref{width})}] is finite even without any high-momentum cutoff.
This is because the dipole radiation already involves a 
high-frequency cutoff \cite{RUO97a} that regularizes small 
$r$ behavior.
The contribution to the optical line shift [Eq.~{(\ref{shift})}]
from the integral (\ref{c}) is not finite. The lower limit of
the integral diverges logarithmically.
Although the radiation kernel (\ref{eq:GDF}) involves a cutoff \cite{RUO97a}, 
the Lamb shift is not treated rigorously. The small-distance
singularities of the dipole radiation kernel may be regularized 
by introducing explicit regularization factors \cite{VRI98}.
However, for the present purposes we may at least obtain
an estimate for the shift by using the cutoff $k_c=1/r_u$ in
the anomalous correlation function (\ref{pairsum})
with the realistic value $r_u=100a_0$ of the triplet $s$-wave
potential~\cite{HOU97}.

\begin{figure}
\begin{center}
%\leavevmode
\epsfig{file=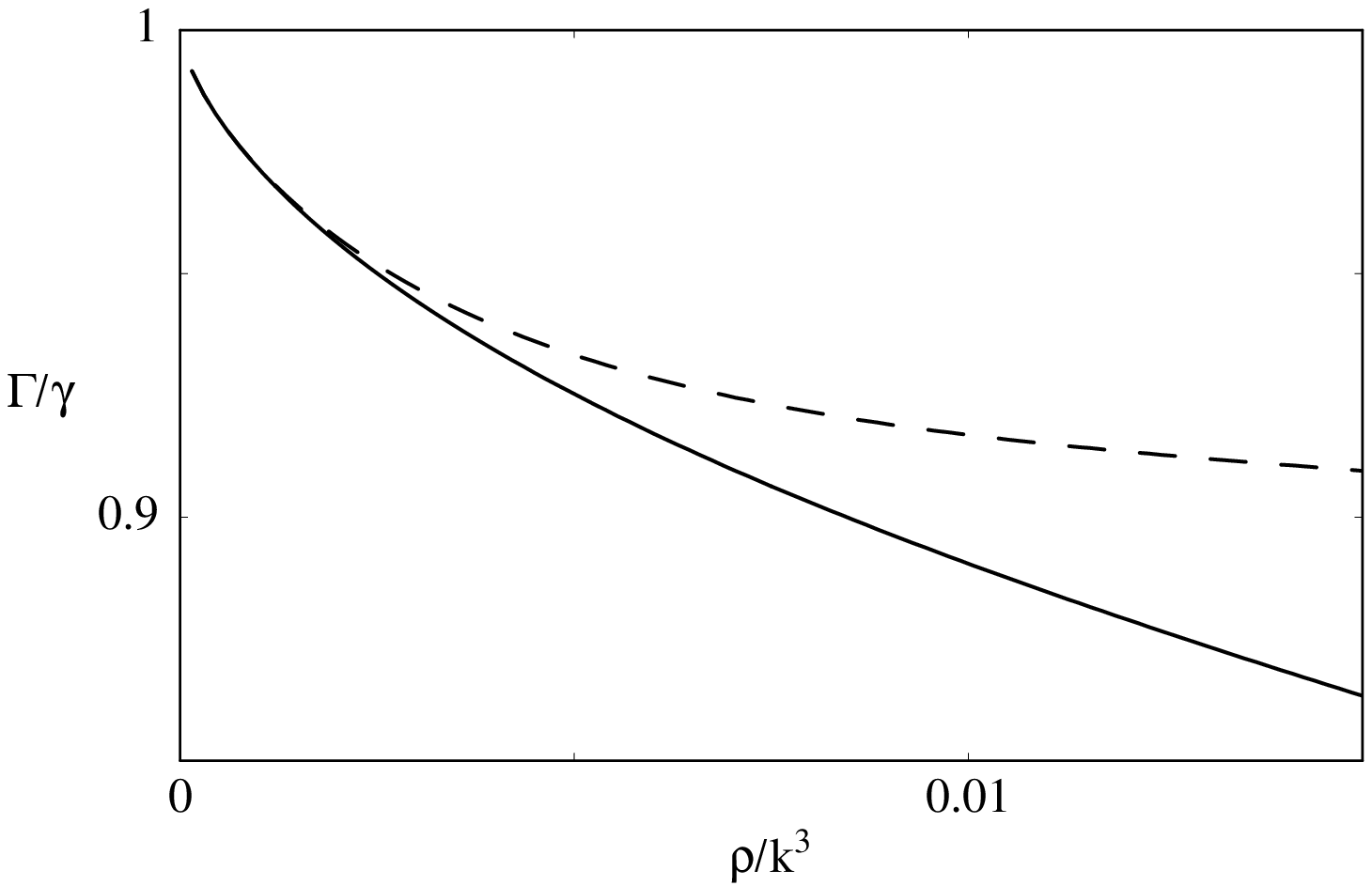,width=7.5cm}
\epsfig{file=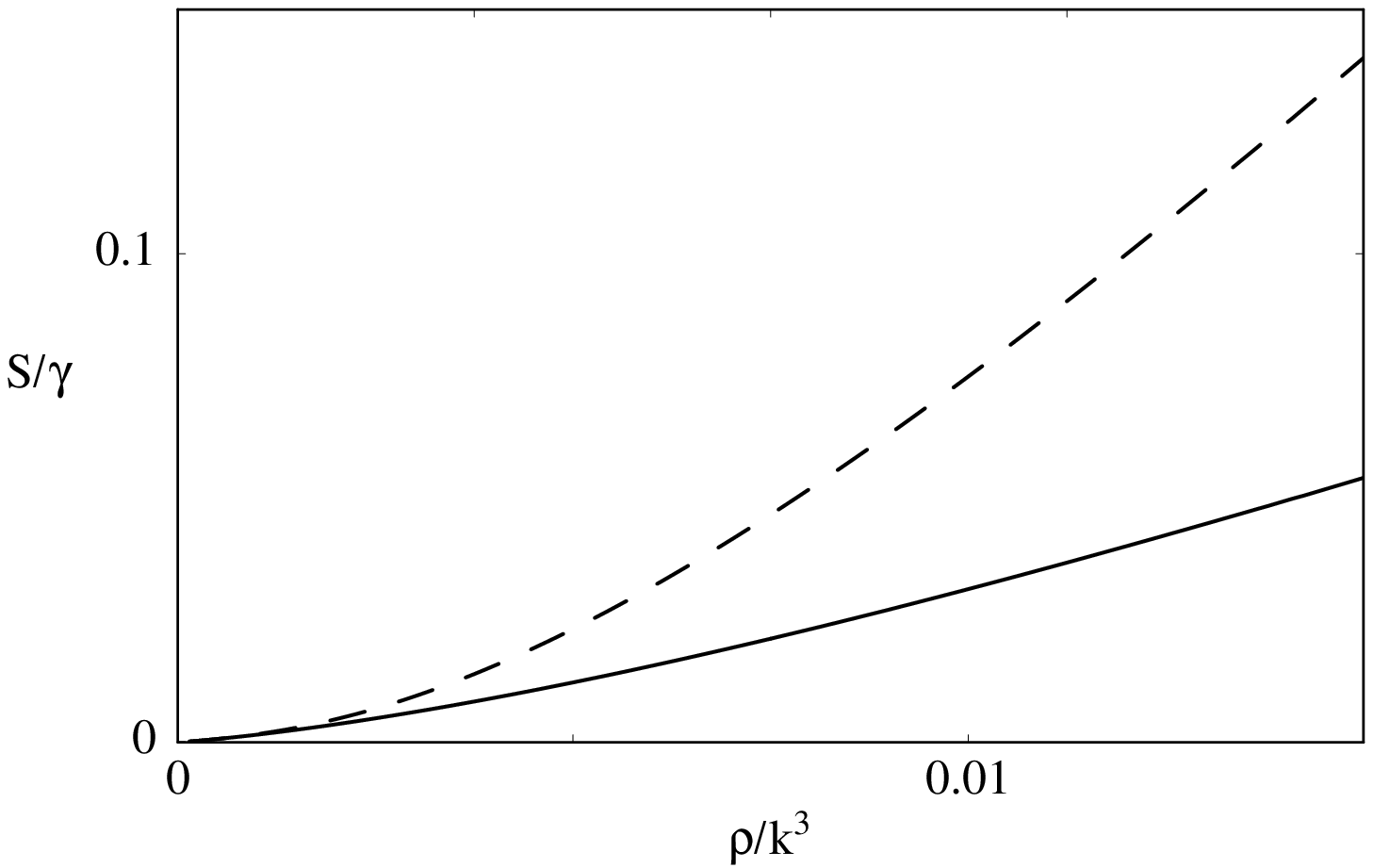,width=7.5cm}
\end{center}
\caption{The optical (a) linewidth and (b) the line shift for the electric
susceptibility, in the absence of the collisional line shift,
as a function of the atom density per cubic optical
wave number of the driving light. The scaling of all the variables is linear.
The solid line represents the optical response in the 
absence of a superfluid state ($\Delta=0$). The BCS pairing broadens 
the resonance line and increses the line shift already at low atom densities.
}
\label{fig:2}
\end{figure}

In Fig. \ref{fig:2} we have shown (a) the absorption linewidth and
(b) the line shift for coherently scattered light from 
Eq.~{(\ref{shift})} without the collisional shift, i.e., by 
assuming ${u}_g={u}_{ge}$ in Eq.~{(\ref{shift})}, 
for $\lambda=900$ nm and for the value of the $s$-wave scattering
length of $^6$Li, $a_g=-2160a_0$ \cite{HOU97}.
In (a) the solid line represents the linewidth 
in the absence of the superfluid state ($\Delta=0$). The line
narrows as a function of the density already at very low densities
\cite{RUO99c}. The presence of the superfluid state broadens 
the optical linewidth (the dashed line).
For the BCS state, even without the collisional shift, 
also (b) the optical line shift is increased. 

It is interesting to emphasize that the optical linewidth
is almost independent of the high-momentum behavior of
the anomalous correlation function (\ref{pairsum}). This
can be seen by introducing a cutoff $k_c$ in Eq.~{(\ref{pairsum})}.
We found \cite{RUO99d} that the optical linewidth is almost 
independent of the cutoff from $k_c=\infty$ to $k_c=1/(500a_0)$ 
indicating that the exact short-range behavior of the
two-body $s$-wave potential is not very crucial for the value of the 
linewidth. Furthermore, the contribution of integral (\ref{c})
in the close neighborhood of the origin to the linewidth is vanishingly
small, and therefore the effects of the BCS pair correlations do not
result from short-range correlations.

\subsection{Summary remarks}
\label{sec:sum}

We calculated the leading quantum-statistical and collisional
corrections to the standard century-old column density result
for the electric susceptibility.
In Eq.~{(\ref{sus})} the corrections to the susceptibility are
encapsulated in the two parameters $\Sigma_1$ and $\Sigma_2$.
Here $\Sigma_2$ represents the collisional line shift
due to the $s$-wave interactions and $\Sigma_1$ position correlations
between different atoms.
The susceptibility was obtained by means of the decorrelation 
approximation {(\ref{p2})} which neglects all the repeated
photon exchange between the same atoms.
These are the microscopic mechanism for the collective optical linewidths
and line shifts \cite{RUO97a}.
Therefore, $\Sigma_1$ is a direct consequence of the quantum-statistical 
correlations; for an uncorrelated atomic sample, with 
$\rho_2(\rv\nu,\rv'\sigma)=\rho_\nu \rho_\sigma$, $\Sigma_1$ vanishes.
In the case of uncorrelated atoms the lowest-order correction to
the optical linewidth results from the collective light scattering
\cite{RUO99c,MOR95,TIG90}. This correction is proportional to atom
density. The quantum statistics is different because the correlation
length itself depends on the density. For instance, in the low-density
limit for an ideal FD gas we obtain \cite{RUO99c}: $\Sigma_1\propto
(\rho/k^3)(\xi_{\up\up}k)\propto \rho^{2/3}/k^2$. Hence, at least in 
the absence of a
superfluid state, Eq.~{(\ref{sus})} not only represents the
lowest-order correction to the susceptibility in terms of microscopic
optical interaction processes between the atoms, but it also correctly
generates the leading low-density correction. 

\section{Bragg spectroscopy}
\label{sec:bragg}

\subsection{Diffraction of atoms}

In this section we consider diffraction of atoms by means of
light-stimulated transitions
of photons between two nonparallel laser beams. When an atomic
beam interacts with a periodic potential formed by a standing
light wave, it can Bragg diffract, analogous to the Bragg diffraction
of x rays from a crystal \cite{KOZ99}. The Bragg diffraction 
has been experimentally used as a technique to split a BE condensate 
\cite{KOZ99} and as a spectroscopic method to probe the density
fluctuations of a BE condensate \cite{STE99,STA99}.

In an $n$th order Bragg scattering process photons are absorbed from one
beam and stimulated to emit into the other $n$ times 
\cite{KOZ99,STE99}. Two different momentum states 
are connected by a $2n$-photon process. The change of the energy 
of a photon upon the scattering satisfies $E=n\hbar\omega$, where $\omega$ 
stands for the frequency difference of the two lasers with wave number $k$.
The fractional change of frequency upon scattering is
assumed to be negligible, $|\omega|\ll kc$, so that
one finds the familiar relation between the change of the wave vector 
$\Dkappav$ and the scattering angle $\theta$,
\beq
|\Dkappav|=2n k \sin(\theta/2)\,.
\label{mome}
\eeq

\subsection{Dynamical structure function}
\label{sec:stru}

In Bragg spectroscopy, the two intersecting laser beams create
a moving standing wave with a periodic intensity modulation
$I(\rv t)=I \cos(\Dkappav\cdot\rv-\omega t)$ \cite{STA99}.
The intensity modulation creates an optical potential $V(\rv t)
\equiv V_0 \cos(\Dkappav\cdot\rv-\omega t)$ which
couples to the local number density of ground-state atoms.
The dependence of $V_0$ on the atomic level scheme and on light 
polarizations is analyzed in Ref.~\cite{JAV95b}.
We consider a situation where the internal sublevel of the 
ground-state atom does not change in the scattering process.
The corresponding Hamiltonian density reads
\beq
\Hc_B= V_0 \cos(\Dkappav\cdot\rv-\omega t)\psi^\dagger_{g\nu}
(\rv)\psi_{g\nu}(\rv)\,.
\label{hb}
\eeq
According to Fermi's golden rule the excitation
rate is then $2\pi/\hbar(V_0/2)^2 S(\Dkappav, \omega)$ \cite{STA99}, where
$S(\Dkappav, \omega)$ is the dynamical structure function  
\cite{For75}
\bea
S(\Dkappav, \omega)&\equiv& {1\over Z}\sum_{i,f} e^{-E_i/k_BT}
|\< i|\hat\rho(\Dkappav)|f\>|^2\nonumber\\
&&\times \delta(\hbar\omega+E_f-E_i)\,.
\label{stru1}
\eea
Here $Z$ denotes the grand partition function and the expectation value
of the density fluctuation operator,
\beq
\hat\rho(\qv)=\sum_\nu\int d^3r\, e^{-i\qv\cdot\rv}\psi^\dagger_{g\nu}
(\rv)\psi_{g\nu}(\rv)\,,
\eeq
is summed over all possible final states $|f\>$, with the energy
$E_f$, and thermally averaged over initial states $|i\>$, with the energy
$E_i$. By using the completeness of $|f\>$ and $\hat\rho(-\qv)=
\hat\rho^\dagger(\qv)$ we may write Eq.~{(\ref{stru1})} as
\bea
S(\Dkappav, \omega)&=& {1\over 2\pi\hbar}\sum_{\nu,\eta}
\int dt\, d^3 r_1 d^3 r_2\,
e^{i\omega t} e^{i\Dkappav\cdot(\rv_1-\rv_2)} \nonumber\\
&&\times \< \psi^\dagger_{g\nu}
(\rv_10)\psi_{g\nu}(\rv_10)\psi^\dagger_{g\eta}(\rv_2t)\psi_{g\eta}
(\rv_2t) \>\,.
\label{stru}
\eea
We define the static structure function by
\beq
\bar{S}(\Dkappav)\equiv \hbar\int^\infty_{-\infty} d\omega\,
S(\Dkappav,\omega)\,.
\label{static}
\eeq

The dynamical structure function mirrors the velocity distribution
of atoms and contains qualitative signatures of BE and FD statistics
\cite{JAV95b}. It displays the modifications of the velocity distribution
due to the quantum statistics including the Fermi {\it inhibition} and
the Bose {\it enhancement} of the scattering process.
For the case of two BE condensates it can also
exhibit a dramatic dependence of the spectrum on the relative phase
between the two condensates \cite{RUO97b,PIT99}, and the Bragg diffraction
could possibly be used as a technique of measuring the relative condensate
phase. This is because, due to the macroscopic quantum coherence of the
BE condensates, the uncertainty in the initial state of the Bragg diffraction
may result in a destructive or constructive interference of the transition
amplitudes. The structure function may also provide information
about the high-energy quasiparticle excitations \cite{CSO98}.
Here we study the qualitative
signatures of a superfluid state in the structure function of a FD gas.

\subsubsection{Ideal Fermi-Dirac gas}

First, we consider an ideal FD gas studied in Ref.~\cite{JAV95b}.
We assume a translationally invariant space. In that case
the correlation function in Eq.~{(\ref{stru})} depends only on 
$\rv\equiv \rv_2-\rv_1$. We are interested in the incoherent
scattering processes corresponding to nonforward directions with
$\Dkappav\neq{\bf 0}$.
In the absence of a superfluid state the correlation function
in Eq.~{(\ref{stru})} with $\nu\neq\eta$ represents only coherent
scattering events. With $\nu=\eta$ we obtain
\bea
\lefteqn{\< \psi^\dagger_{g\nu}
(00)\psi_{g\nu}(00)\psi^\dagger_{g\eta}(\rv t)\psi_{g\eta}
(\rv t) \>}\nonumber\\
&&=\rho^2+\< \psi^\dagger_{g\nu}(00)\psi_{g\eta}(\rv t)
\> \< \psi_{g\nu}(00) \psi^\dagger_{g\eta}(\rv t)\>\,.
\label{braggcor}
\eea
For $\Dkappav\neq{\bf 0}$ we obtain the dynamical structure
function from Eq.~{(\ref{stru})}
\beq
S = {1\over\hbar}\sum_{{\bf k},\nu} 
\delta\left(\omega+{\omega_R}-{\hbar{\bf k}\cdot\Dkappav\over m}\right)
{\bar{n}}_{\kv,\nu}(1-{\bar{n}}_{\kv-\Dkappav ,\nu})\,.
\label{simstru}
\eeq
Here $\bar{n}_{\kv,\nu}\equiv [\exp{(\epsilon_{\kv} /k_BT)}/z+1]^{-1}$ 
denotes the FD occupation numbers and $z$ fugacity. We have also
defined the effective recoil frequency $\omega_R$ by
\begin{equation}
{\omega_R} = {\hbar |\Dkappav|^2\over 2 m}\,.
\label{eq:omega}
\end{equation}

Expression {(\ref{simstru})} describes a scattering process 
in which an atom in the
ground state $\nu$ with the c.m. state $\kv$ scatters to the c.m. state
$\kv-\Dkappav$ still remaining in the state $\nu$. 
The delta function dictates the energy conservation, which coincides with
the theory for Doppler velocimetry of atoms \cite{hetdyn} shifted by 
the effective recoil frequency $\omega_{R}$~\cite{JAV95b}. 

Classical atoms obey Maxwell-Boltzmann statistics and their velocities
are normally distributed resulting in a Gaussian-shaped dynamical
structure function \cite{JAV95b}. Firstly, FD statistics modifies the
velocity distribution; even an ideal FD gas at $T=0$ exhibits a finite
width in Eq.~{(\ref{simstru})}. Secondly, the quantum degeneracy 
affects the scattering processes.
The product of the occupation numbers in Eq.~{(\ref{simstru})} 
indicates the Fermi {\it inhibition}: The scattering events in which
an atom recoils to an already occupied state are forbidden by
the Pauli exclusion principle.

It is illustrative to describe the Fermi inhibition in momentum space
\cite{RUO99c}. At $T=0$ the fermionic atoms fill the Fermi sphere
with $\bar{n}_{\kv,\nu}=\Theta(k_F-|\kv|)$. 
The scattering satisfies Eq.~{(\ref{mome})}. For the first-order Bragg
diffraction, with $n=1$, all atoms are scattered out of the Fermi sea,  
if $|\Dkappav| > 2k_F$. Moreover, for small scattering angles,
\beq
\sin(\theta/2)< {k_F\over|\Dkappav|}= \({6\pi^2\rho\over
|\Dkappav|^3}\)^{1/3}\,, 
\eeq 
the incoherent atomic recoil events are forbidden. When the density is
increased, at $k_F\geq |\Dkappav|$ the scattering is at least partially
suppressed to all nonforward directions.

We consider a situation where the densities of the internal sublevels
are equal $\rho\equiv\rho_\up=\rho_\down\equiv N/V$. Here $N$ denotes
the number of atoms.
The dynamical structure function for an ideal FD gas was displayed
in Ref.~\cite{JAV95b}. At $T=0$, and in the degenerate regime with
$|\Dkappav| < 2k_F$, the result exhibits a characteristic shape of a
wedge consisting of a linear and a quadratic part. The static structure
function per total number of atoms may be evaluated from 
Eq.~{(\ref{simstru})}. For $|\Dkappav|> 2k_F$ we obtain
$\bar{S}(\Dkappav)/(2N)=1$ and for $0<|\Dkappav| < 2k_F$
\beq
{\bar{S}(\Dkappav)\over 2N}={1\over16 k_F |\Dkappav|}
\( 12 |\Dkappav|^2-{|\Dkappav|^4\over k_F^2}\)\,.
\label{statabs}
\eeq
For free atoms $\bar{S}(\Dkappav)/(2N)=1$ and Eq.~(\ref{statabs})
describes the inhibited scattering.

In the previous discussion we ignored the Rabi oscillation dynamics
and considered only the transition rates. This is a good approximation
when only a small fraction of atoms is scattered, i.e., when the
coupling time $t_0$ is much shorter than the oscillation period 
$t_0\ll 1/\Omega$ \cite{RUO97b}. Here $\Omega$ denotes the two-photon
Rabi frequency. To observe the qualitative features in the spectrum
the coupling time should be at least of the order of the characteristic time 
scale in the spectrum, which according to Ref.~\cite{JAV95b} is 
$(\omega_R\epsilon_F/\hbar)^{-1/2}$.

\subsubsection{BCS pairing}

In the presence of a superfluid state the correlation function in
Eq.~{(\ref{stru})} at $T=0$ reads (for $\nu\neq\eta$):
\bea
\lefteqn{
\< \psi^\dagger_{g\nu}
(00)\psi_{g\nu}(00)\psi^\dagger_{g\eta}(\rv t)\psi_{g\eta}
(\rv t) \>}\nonumber\\
&&=\rho^2+|\< \psi_{g\nu}(00)\psi_{g\eta}(\rv t)
\>|^2\,.
\label{braggcor2}
\eea
For the case $\nu=\eta$ we obtain Eq.~{(\ref{braggcor})}.
Analogously to Eqs.~{(\ref{gaus})} we need to evaluate 
Eqs.~{(\ref{braggcor})} and~{(\ref{braggcor2})} by means of the
Bogoliubov quasiparticles [Eq.~{(\ref{bogo})}]. The dynamical 
structure function for incoherent scattering ($\Dkappav\neq{\bf 0}$) 
in the quasiparticle vacuum is
\bea
\lefteqn{S(\Dkappav,\omega)={1\over4}\sum_\kv 
\delta(\hbar\omega+E_\kv+E_{\kv-\Dkappav})}\nonumber\\
&&\mbox{}\times\[ {\Delta^2\over
E_\kv E_{\kv-\Dkappav}}+\(1-{\xi_\kv\over E_\kv}\)
\(1+{\xi_{\kv-\Dkappav}\over E_{\kv-\Dkappav}}\)\]\,,
% &&\times{1\over4}\,\delta(\hbar\omega+E_\kv+E_{\kv-\Dkappav})\,,
\label{bcsdynamic}
\eea
where the quasiparticle energies $E_\kv$ are defined in Sec.~\ref{sec:bcs}.
Expression (\ref{bcsdynamic}) describes creations of pairs of quasiparticles
separated by the wave vector $\Dkappav$.
We note that for a superfluid state in the homogeneous space there exists 
a finite energy gap in the excitation spectrum 
$-\hbar\omega=E_\kv+E_{\kv-\Dkappav}\geq 2|\Delta|$.
The corresponding expression for the static structure function 
$\bar{S}(\Dkappav)$ may be obtained
from Eqs.~{(\ref{bcsdynamic})} and~{(\ref{static})}. 
%\beq
%\bar{S}(\Dkappav)={1\over4}\sum_\kv \[ {\Delta^2\over
%E_\kv E_{\kv-\Dkappav}}+\(1-{\xi_\kv\over E_\kv}\)
%\(1+{\xi_{\kv-\Dkappav}\over E_{\kv-\Dkappav}}\)\]\,.
%\label{bcsstatic}
%\eeq

\begin{figure}
\begin{center}
%\leavevmode
\epsfig{file=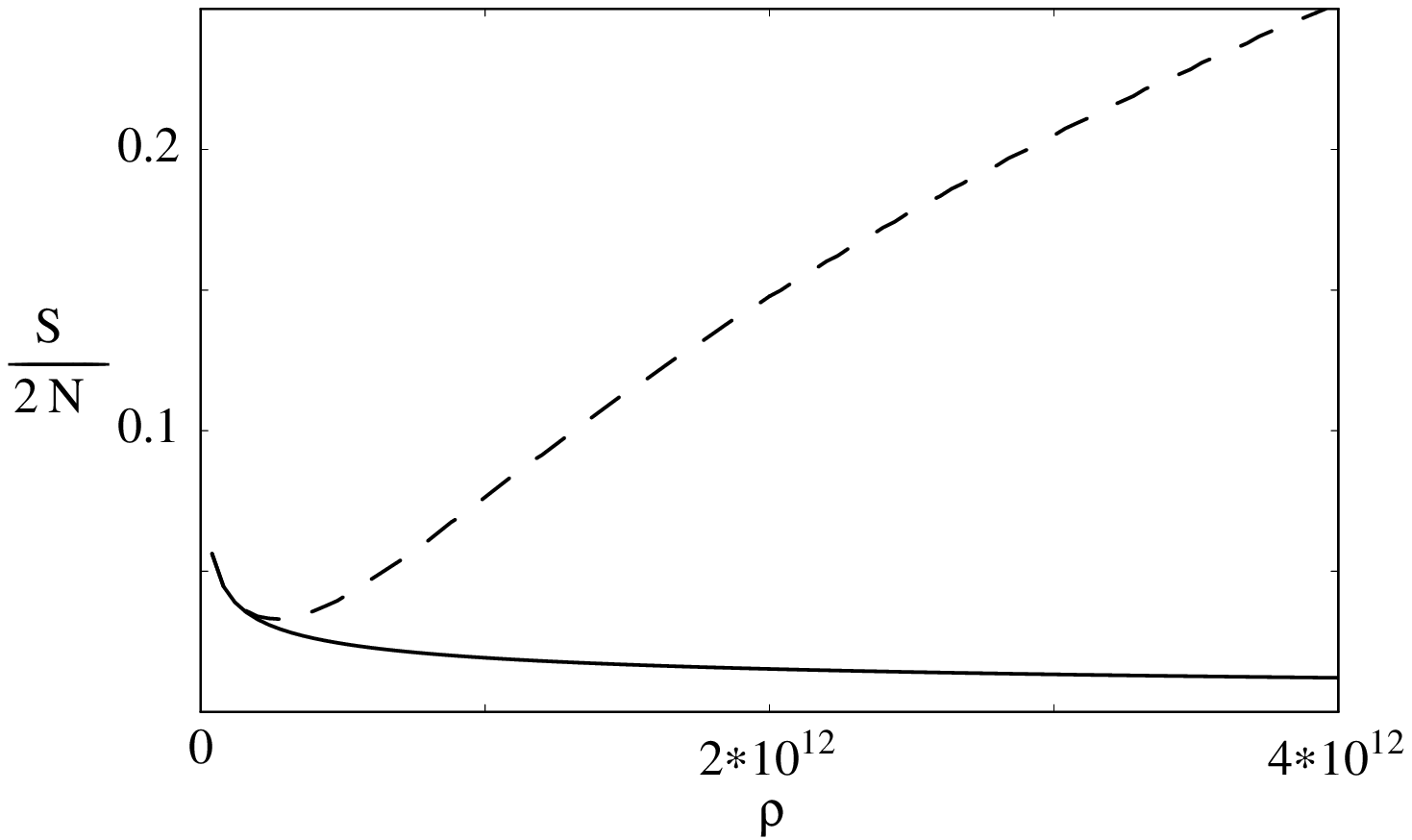,width=7.5cm}
\epsfig{file=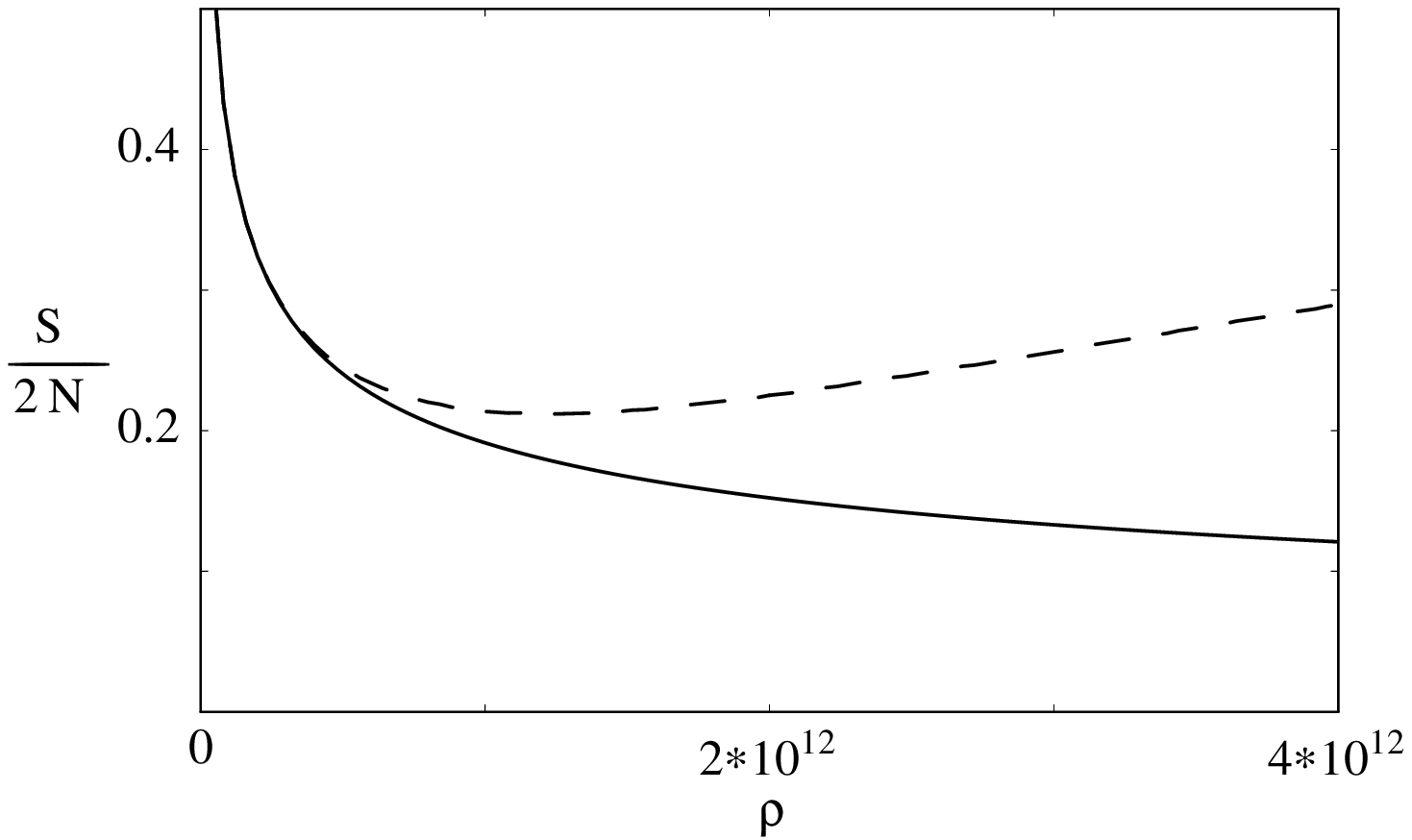,width=7.5cm}
\epsfig{file=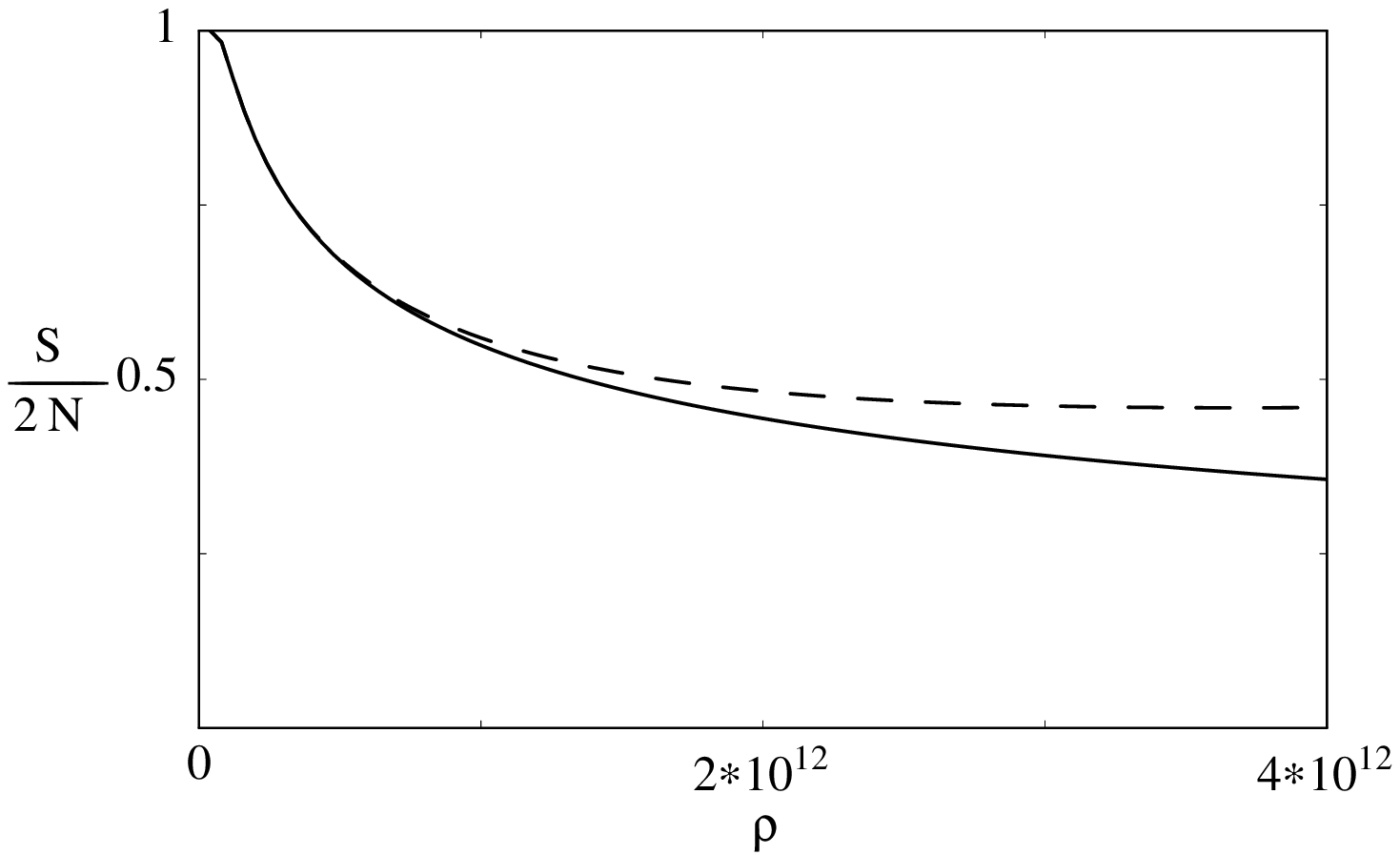,width=7.5cm}
\end{center}
\caption{
The static structure function per the total number of atoms
$\bar{S}(\Dkappav)/(2N)$ as a function of the atom density $\rho$ in
units of cm$^{-3}$. The change of the atomic 
recoil wave number upon scattering is (a) $10^3$ cm$^{-1}$, (b) $10^4$ 
cm$^{-1}$, and (c) $3\times 10^4$ cm$^{-1}$.
The solid line represents the diffraction in the 
absence of a superfluid state ($\Delta=0$). The BCS pairing dramatically
increases the incoherent nearforward scattering
already at low atom densities.
}
\label{fig3}
\end{figure}

In Fig.~\ref{fig3} we show the static structure function 
$\bar{S}(\Dkappav)/(2N)$ for a FD gas
as a function of the density $\rho$ for three characteristic values
of $|\Dkappav|$. The $s$-wave scattering length $a_g=-2160a_0$.
The solid line represents an ideal FD gas in the
absence of a superfluid state. The superfluid state dramatically
increases the structure function  (the dashed line) 
for nearforward scattering. The BCS pairing mixes particles and
holes near the Fermi surface increasing the number of available
scattering channels. This effect is particularly striking for
the case of small recoil momentum corresponding to nearforward
scattering.

We may also consider situations where the internal state of 
atoms is changed in the scattering process. In this case the two
ground states $|g,\up\>$ and $|g,\down\>$ are coupled through a
common excited state by the intersecting laser beams. For instance,
the scattering rate for the transition $|g,\down\>\rightarrow|g,\up\>$ 
is proportional to $\< \psi^\dagger_{g\down}
(00)\psi_{g\up}(00)\psi^\dagger_{g\up}(\rv t)\psi_{g\down}
(\rv t) \>$ and depends on the quasiparticle pairing.

\section{Conclusions}
\label{sec:con}

We studied the interaction of light with a two-species 
atomic superfluid gas. Firstly, we considered the propagation
of light and evaluated the quantum-statistical corrections to
the standard column density result for the electric susceptibility.
Secondly, we analyzed the Bragg diffraction of atoms by means
of light-stimulated transitions of photons between two laser beams.
The effects of BCS pairing may be understood in terms of enhanced
incoherent scattering processes resulting in the increased 
optical linewidth, line shift, and static structure function.
These optical properties could possibly signal the presence 
of the superfluid state and determine the value of the BCS order 
parameter in dilute atomic FD gases.

One particularly promising candidate to
undergo the BCS transition and to become a superfluid is spin-polarized
atomic $^6$Li. Atoms in two different internal levels can interact 
via $s$-wave scattering and the $^6$Li atom has an anomalously 
large and negative $s$-wave scattering length $a\simeq-2160a_0$.
The hyperfine states $|m_s=1/2,m_i=1\>$ and $|m_s=1/2,m_i=0\>$
of $^6$Li have been 
predicted to undergo a superfluid transition at $10^{-8}$ K with 
a density of $10^{12}$ cm$^{-3}$ \cite{STO96,HOU97}.
Here $m_s$ and $m_i$ denote the electron and the nuclear spin
components.

We assumed a translationally invariant system.
A FD gas in a harmonic trap
may be considered locally homogeneous \cite{HOU97}, provided that
the trap length scale $l=(\hbar/m\omega)^{1/2}$ is much larger than
the correlation lengths, $\xi_{\up\up}$ and $\xi_{\up\down}$.
The spatial confinement introduces an uncertainty in the
recoil momentum. In the case of Bragg spectroscopy, the coherent scattering 
is negligible, if the change of the wave number
of the atoms upon scattering is larger than the inverse size scale
of the atomic sample $1/l\lesssim |\Dkappav|$.

\subsection*{Acknowledgements}

We acknowledge discussions with H. T. C. Stoof and P. Zoller.
This work was financially supported by EC through the TMR 
Network ERBFMRXCT96-0066.

\end{document}